\documentclass[11pt]{article}
\usepackage{amsfonts}
\usepackage{amsthm}
 \usepackage{bezier}

\usepackage{epic,eepic,amsmath,amssymb,color,graphicx}

\def\be{\begin{equation}}
\def\ee{\end{equation}}
\def\bea{\begin{eqnarray}}
\def\eea{\end{eqnarray}}

\def\be{\begin{equation}}
\def\ee{\end{equation}}
\def\bea{\begin{eqnarray}}
\def\eea{\end{eqnarray}}

\def\1{\'{\i}}                           

\def\bq{\mathbf{q}}
\def\bp{\mathbf{p}}
\def\cM{{\mathcal M}}
\def\RR{\mathbb{R}}
\def\dd{{\rm d}}

\newcommand\Om\Omega

\newcommand{\bL}{\mathbf{L}}

\def\>#1{{\mathbf#1}}

  \def\la{{\lambda}}
 \def\ga{{\gamma}}
 \def\del{{\delta}}
 \def\bt{{\beta}}
 
 \def\xxi{{\xi}}
 \def\ch{{\chi}}
\def\k{{\kappa}}

\newcommand{\Kep}{_{\mathrm{KC}}}
\newcommand{\Harm}{_{\mathrm{O}}}

\def\ctea{{A}}
\def\cteb{{B}}

\newcommand{\hbq}{\hat{\mathbf{q}}}
\newcommand{\hbp}{\hat{\mathbf{p}}}

\newcommand{\hH}{\hat{\cal{H}}}
 
\newcommand{\hq}{\hat{q}}
\newcommand{\hp}{\hat{p}}
\def\rmi{{\rm i}}

\begin{document}

\thispagestyle{empty}

  \noindent
 {\Large{\bf {Superintegrable quantum oscillator and Kepler-Coulomb\\[6pt] systems on curved spaces }}}

\medskip 
\medskip 
\medskip

\begin{center}
{\sc \'Angel Ballesteros$^a$, Alberto Enciso$^b$, Francisco J. Herranz$^{a,}$\footnote{
 Based on the contribution presented at ``The XXIX International Colloquium on Group-Theoretical Methods in Physics",
August 20-26, 2012, 
Chern Institute of Mathematics,
 Tianjin, China.\\[4pt]
\indent \indent {\em Proceedings of the XXIX International Colloquium on Group-Theoretical Methods in Physics}, Nankai Series in Pure, Applied Mathematics and Theoretical Physics, Chengming Bai, Jean-Pierre Gazeau and Mo-Lin Ge (eds.), World Scientific: Singapore, 2013, to appear 
},\\[4pt] Orlando Ragnisco$^{c}$ and Danilo Riglioni$^{d}$}
\end{center}
\medskip 
\medskip 

\noindent
$^a$ Departamento de F\'\i sica,  Universidad de Burgos,
E-09001 Burgos, Spain\\
angelb@ubu.es,\quad fjherranz@ubu.es  \\[4pt] 
 $^b$ Instituto de Ciencias Matem\'aticas,  CSIC, E-28049 Madrid,
Spain\\
aenciso@icmat.es \\[4pt] 
 $^c$ Dipartimento di Fisica,   Universit\`a di Roma Tre and Instituto
Nazionale di Fisica Nucleare sezione di Roma Tre,  
I-00146 Roma, Italy\\
ragnisco@fis.uniroma3.it  \\[4pt] 
 $^d$ Centre de Recherches Math\'ematiques,  Universit\'e de Montr\'eal,   H3C 3J7, Canada\\
 riglioni@CRM.UMontreal.ca

  \medskip 
\bigskip
\bigskip

\begin{abstract} 
\noindent
 An overview of  maximally superintegrable classical Hamitonians on spherically symmetric spaces is presented. It turns out that each of these systems can be considered either as an oscillator or as a Kepler--Coulomb Hamiltonian. We show that two possible quantization prescriptions for all these curved systems arise if we impose that superintegrability is preserved after quantization, and we prove that both possibilities are gauge equivalent.  
\end{abstract} 

\vfil

\newpage


\section{Bertrand spacetimes}

  Bertrand's theorem asserts that any  three-dimensional (3D) {spherically symmetric natural Hamiltonian} system $$H=\frac12 \bp^2+ V(|\bq|)$$   that has a stable circular trajectory  passing through each point of $\RR^3$   and all whose bounded trajectories are closed is  
either a harmonic oscillator,  $V(r)=\omega^2\bq^2+a$, or 
 a  Kepler--Coulomb (KC) system, $V(r)=\frac{k}{|\bq|}+b$.
  An extension of such a classical    theorem was found by Perlick~\cite{Perlick} in a  General Relativity framework.     Consider a $(3+1)$D  spherically symmetric  spacetime $(\cM\times\RR,\eta)$, where $\cM$ is a   $3$-manifold. Then the   Lorentzian metric  $\eta$ can be written as
$$
\eta=h(r')^2\,\dd r'^2+r'^2\big(\dd\theta^2+\sin^2\theta\,\dd\varphi^2\big)-\frac{\dd t^2}{V(r')} = g-\frac{\dd t^2}{V(r')} ,
\label{za}
$$
where $g$  is a Riemannian metric on $\cM$ and $h(r'),V(r')$ are smooth functions.  
  A  {\em   trajectory} in the   spacetime is the projection of an inextendible timelike geodesic to a  constant time leaf $\cM\{t_0\}$. 
  The Lorentzian (3+1)-manifold $(\cM\times \RR,\eta)$ is a  {\em Bertrand spacetime} if it is verified that: (i)  there is a circular ($r'=\text{constant}$) trajectory passing through each point of $\cM$, and (ii) 
such circular trajectories are stable. Under these assumptions, 
  Perlick undertook the classification of all     Bertrand spacetimes    finding   two multiparametric families of metrics by  obtaining  explicitly the functions $h(r'),V(r')$.
  
   These results  can be extended to arbitrary dimension   and expressed in a conformally flat form. 
 Let us 
consider an $(N+1)$D spherically symmetric  spacetime $(\cM\times\RR,\eta)$, where hereafter $\cM$ is  an $N$-manifold.  The Lorentzian metric $\eta$ can be written as
\be
\eta=  g-\frac{\dd t^2}{V(r)},\qquad g=   f(|\bq|)^2\,\dd\bq^2 = f(r)^2(\dd r^2+r^2\dd\Om^2) ,
 \label{aax}
\ee
where
 $\bq=(q_1,\dots,q_N)$,  $\dd\bq^2 =\sum_{i=1}^N\dd q_i^2$,   $r=|\bq| = \sqrt{\bq^2}$ and    $\dd\Om^2$  is the standard metric on the unit $(N-1)$D sphere. 
 Hence, $g$ 
defines a  Riemannian metric on $\cM$ with $f(r)$ playing the role of {\em conformal factor} of the Euclidean metric $g_0=\dd\bq^2$.
  The  scalar curvature  of $g$  is generally nonconstant and turns out to be
\be
R=-(N-1)\left( \frac{    (N-4)f'(r)^2+  f(r)  \left(    2f''(r)+2(N-1)r^{-1}f'(r)  \right)}   {f(r)^4  } \right) .
\label{aac}
\ee

By  starting from  Perlick's classification of  $(3+1)$D Bertrand spacetimes, if we  change the initial  radial coordinate in the form $r' ={\cal F}(r)$ and we accordingly transform the initial  Perlick's parameters,  then it can be proven through  very cumbersome computations that the metric $\eta$ of an $(N+1)$D Bertrand spacetime belongs to one of the following classes~\cite{metrics}:

\noindent
$\bullet$  {\bf Type I}. This is    a   {\em three-parametric}  family of metrics depending on $(\bt;\k,\xxi)$ where   $\bt$ is a positive rational  number and  $\k,\xxi$ are real constants:
\be
\eta_{\rm I}=\frac{1}{r^2\left( r^{-\bt}+\k r^\bt  \right)^2}\,\dd\>q^2-\frac{\dd t^2}{  \left(r^{-\bt}-\k r^\bt \right)+\xxi} .
\label{zaf}
\ee

\noindent
$\bullet$  {\bf  Type II}. This is a  {\em four-parametric} family  depending on $(\ga;\la,\del,\ch)$ where   $\ga$  is a positive rational number and  $\la^2,\del,\ch$ are real constants:
\be
\eta_{\rm II}=\frac{\left( r^{-2\ga}+\la^2 r^{2\ga}-2\del\right)}{r^2\left( r^{-2\ga}-\la^2 r^{2\ga}  \right)^2}\,\dd\>q^2-
\frac{\dd t^2}{ \left(r^{-2\ga}+\la^2 r^{2\ga}-2\del \right)^{-1}+\ch }  .
\label{zag}
\ee

The sets of parameters  $(\bt;\k,\xxi)$ and $(\ga;\la,\del,\ch)$ can  be, in fact, regarded as  {\em deformation parameters} of  the {\em flat} Minkowskian metric, since this  is recovered by  a contraction limit of (\ref{zaf}) and (\ref{zag})  as follows:
$$
 \begin{array}{llllllll}
\eta_{\rm I}&=\dd\>q^2- \dd t^2  &\ \  \mbox{for}\ \  \bt&=1,\quad \k&=0,\qquad\  \  \xxi&=1/\varepsilon^2;\quad t&\to t/\varepsilon,\quad \varepsilon &\to 0. \cr
 \eta_{\rm{ II}}&=\dd\>q^2- \dd t^2 &\ \ \mbox{for}\ \   \ga&=1,\quad \la&=\del=0,\  \ \ch&=1/\varepsilon^2;\quad t&\to t/\varepsilon,\quad \varepsilon &\to 0.
\end{array}
$$


\section{Classical Bertrand Hamiltonians}

Timelike geodesics in  an $(N+1)$D  spacetime with metric $\eta$ (\ref{aax}) are naturally related to   trajectories of the $N$D classical Hamiltonian  in $\cM$ given by
\be
  H= \frac{\bp^2}{2f(|\bq|)^2}+V(|\bq|)=\frac{p_r^2+r^{-2}\bL^2}{2f(r)^2}  +V(r),
\label{zg}
\ee
where $\>p=(p_1,\dots,p_N)$ and $p_r$ 
 are, in this order, the conjugate momenta of the coordinates $\>q$ and  $r$,  
while $\bL$ is the   total   angular momentum.   

Thus, Perlick's study of Bertrand spacetimes provides a complete classification of the central potentials $V(r)$ and spherically symmetric metrics for which the statement of the classical Bertrand theorem remains true. Moreover, we showed that these Hamiltonian systems are still superintegrable~\cite{CMP}.
 Therefore, by considering the classification of  Bertrand spacetimes (\ref{zaf}) and (\ref{zag}),   we write the corresponding {\em classical  Bertrand Hamiltonians} which
are cast into the following two families~\cite{metrics}:

\noindent
$\bullet$  {\bf Type I}.   {\em  Two-parametric} family of Hamiltonians depending on $(\bt;\k)$: 
\be
H_{\rm I}=\frac{1}2  {r^2\left( r^{-\bt}+\k r^\bt  \right)^2}\, \>p^2+A {  \left(r^{-\bt}-\k r^\bt \right)} .
\label{zm}
\ee

\noindent
$\bullet$  {\bf Type II}.     {\em Three-parametric} family   depending on $(\ga;\la,\del)$:
\be
H_{\rm II}=\frac{r^2\left( r^{-2\ga}-\la^2 r^{2\ga}  \right)^2}{2\left( r^{-2\ga}+\la^2 r^{2\ga}-2\del\right)} \, \>p^2 +
\frac{B} { \left(r^{-2\ga}+\la^2 r^{2\ga}-2\del \right)} .
\label{zn}
\ee

We remark that   the    parameters $\xxi,\ch$, which are rather relevant in the metrics,    become  additive constants  in the Hamiltonians. 
The coupling constants  $A,B$ of the potentials  do not appear in the metrics since they can be set equal to 1   through a time scaling. Note also that Bertrand Hamiltonians are, as expected, generalizations  (deformations) of the flat harmonic oscillator and KC systems, which  are recovered  as   particular cases:
 $$
 \begin{array}{lllll}
H_{\rm I}&= \tfrac 12  \>p^2 +A/r&\quad\mbox{for}\quad \bt&=1,\quad \k&=0  . \cr
H_{\rm II}&=\tfrac 12  \>p^2 +B r^2&\quad \mbox{for}\quad  \ga&=1,\quad \la&=\del=0 .
\end{array}
$$

The main properties of   Bertrand Hamiltonians are the following:

 \noindent
{(i)}  {\em Superintegrability}. The  spherical symmetry of any classical Hamiltonian of the type~(\ref{zg}) 
provides  $2N-3$ integrals of motion which are quadratic in the momenta.
{\em Maximally superintegrable} systems are distinguished cases for which an additional constant of the motion exists (which is a component of a Runge--Lenz $N$-vector),  so that $H$ is endowed with  $2N-2$ functionally independent integrals. This has been explicitly proved  for 3D  Bertrand Hamiltonians~\cite{CMP},  finding that the integrals of motion are not, in general, quadratic in the momenta 
(such higher order  is determined by  $\beta,\ga$).

\medskip

\noindent
{(ii)}  {\em Geometric interpretation}.   The 
{\em  intrinsic KC and  oscillator  potentials} on  the $N$D space $\cM$ are defined  by means of  the   radial symmetric Green function $U(r)$   of the Riemannian metric $g$   (\ref{aax})  as
\begin{equation}
U(r) =\int^r\frac{\dd r}{r^2f(r)} ,\qquad  {{\cal U}\Kep(r) :=A\,U (r) },\qquad  {{\cal U}\Harm(r) :=\frac \cteb{U(r)^2} }.
\nonumber
\end{equation}
 If we apply these definitions to the two families of Bertrand Hamiltonians it can be shown~\cite{CQG} that
   {type I}  (\ref{zm})  and  {type II} Hamiltonians  (\ref{zn})    always define, in this order,   intrinsic KC systems 
   and oscillator potentials      on curved spaces.
Consequently,  the extension of the classical Bertrand's theorem to spherically symmetric spaces only provides either oscillators or KC systems.
\medskip

\noindent
{(iii)}  {\em St\"ackel transform}. Both types of  Bertrand Hamiltonians are 
 equivalent via the St\"ackel transform~\cite{Millera}. If we denote (\ref{zm}) as
 $H_{\rm I}=T_{\rm I}+V_{\rm I}$ and consider $H_{0,\rm I}= T_{\rm I}+B$ and $U_{\rm I}= V_{\rm I}/A+C$, then $H_{0,\rm I}/U_{\rm I}\equiv  H_{\rm II}$ 
provided that
$$
\bt=2\ga,\quad \k=-\la^2,\quad C=-2\del.
\label{zo}
$$
Conversely,  Hamiltonians of   type I can be obtained from those of  type II,   so establishing  a   {\em  ``duality"} between curved  oscillators and KC systems~\cite{metrics}.

\medskip

We illustrate the above results by displaying in Table 1 some relevant specific Bertrand Hamiltonians that arise for $\bt=1,2\leftrightarrow\ga=\frac 12,1$;  all of them have quadratic integrals~\cite{metrics}.  The remaining   parameters, $\k$ and $(\la,\del)$ are related with the  curvature of the space. In each row we present a KC system  with its  St\"ackel  equivalent  (``dual") oscillator. The names ``Euclidean", ``Taub--NUT"\dots, refer to the underlying Riemannian space. Notice that the KC system on the {\em three}   spaces of {\em constant} sectional curvature $\k$ ($>,<,=0$ for the sphere, hyperbolic space, Euclidean space, respectively) is recovered  in the cases 1A and 1B of   type I $(\bt=1)$, meanwhile their corresponding oscillator potential appears  in the cases 2A.1 and 2B.1 of   type II $(\ga=1)$,  with $\la=\del$  now playing the role of the constant  curvature.

 \begin{table}[t] { 

 \noindent
\caption{  
    Examples of $N$-dimensional classical Bertrand Hamiltonians. }

\vspace{2ex}
\label{table1}

\noindent  \begin{tabular}{ll}
  
      \hline
   &   \\[-8pt]
 Type I: KC Hamiltonians $(\bt;\k)$&\quad   Type II: Oscillator Hamiltonians $(\ga;\la,\del)$\\[6pt]
  \hline
   &   \\[-8pt]
  $\bullet$ 1A  $(\bt=1; \k=0)$    Euclidean &\quad  {$\bullet$
1A $(\ga=\frac 12 ;\la=0,\del)$  Taub--NUT}\\[4pt]
$\displaystyle{ H=\frac 12 \>p^2+ \frac{\ctea}{r} }$  &\quad  $\displaystyle{ H=  \frac{ r}{2(1-2\del r)} \,\>p^2+  \frac{\cteb r}{(1-2\del r)}   } $\\[16pt]
  $\bullet$ 1B  $(\bt=1;\k)$   Sphere/Hyperbolic&\quad  {$\bullet$
1B  $(\ga=\frac 12 ;\la,\del )$   Darboux IV}\\[4pt]
$\displaystyle{ H=\frac 12(1+\k r^2)^2 \>p^2+\ctea\, \frac{(1-\k r^2)}{r} }$ &\quad 
$ \displaystyle{ H=\frac { (1-\la^2 r^2)^2r}{2(1+\la^2 r^2-2\del r)}\,\>p^2+\frac{\cteb r}{(1+\la^2 r^2-2\del r)}
 }$\\[12pt]
  \hline
   &   \\[-8pt]
   {$\bullet$ 2A  $(\bt=2;\k=0)$ }&\quad  {$\bullet$
2A.1 $(\ga=1;\la=0,\del=0)$   Euclidean}\\[4pt]
$\displaystyle{ H=\frac  { \>p^2}{2r^2}+\frac{\ctea }{r^2} }$&\quad 
 $\displaystyle{ H=\frac 12   \>p^2 + {\cteb r^2}} $\\[16pt]
 &\quad\mbox {$\bullet$
2A.2  $(\ga=1;\la=0,\del)$ Darboux III}\\[4pt]
 &\quad 
$ \displaystyle{ H=\frac {\>p^2}{2(1-2\del r^2)} +\frac {\cteb r^2}{ (1-2\del r^2)}} $\\[14pt]
    {$\bullet$ 2B  $(\bt=2;\k)$ } &\quad  {$\bullet$
2B.1  $(\ga=1;\la,\del=\la)$  Sphere/Hyperbolic}\\[4pt]
$ \displaystyle{ H=\frac  { (1+\k  r^4)^2}{2r^2}\, \>p^2 +\ctea\, \frac{ (1-\k  r^4) }{ r^2} }$ &\quad 
$ \displaystyle{ H=\frac 12 {(1+\la r^2)^2}{\>p^2} +\frac {\cteb r^2}{ (1-\la r^2)^2}} $\\[14pt]
&\quad\mbox {$\bullet$
2B.2  $(\ga=1;\la,\del)$ }\\
 &\quad 
$ \displaystyle{  H=\frac  { (1-\la^2 r^4)^2}{2(1+\la^2 r^4-2\del r^2)}\,\>p^2+\frac{\cteb r^2 }{(1+\la^2 r^4-2\del r^2)} } $\\[12pt] 

  \hline
   \end{tabular}
} 
\end{table}


\section{Quantum Bertrand Hamiltonians}

In order to obtain the quantization of the classical  Bertrand Hamiltonians (\ref{zm}) and (\ref{zn}), the ordering ambiguity coming from the kinetic term can be fixed by imposing the resulting quantum Hamiltonians ${\hH}$ to be maximally superintegrable. This means that the classical integrals are promoted to $(2N-2)$ algebraically independent  operators that commute with ${\hH}$.  
We consider 
the  quantum position and momenta operators, $\hbq$, $\hbp$, such that 
$[\hq_i,\hp_j]=\rmi \hbar \delta_{ij}$, $\hq_i=q_i$,  $\hp_i=-\rmi \hbar\frac{\partial}{\partial q_i}$, and we
 present  two  quantization prescriptions that fulfil  the superintegrability condition~\cite{annals}: the so-called Schr\"odinger quantization and the conformal Laplace--Beltrami (LB) one.


\noindent
$\bullet $  {\em  ``Direct"  or Schr\"odinger quantization}~\cite{Iwai}. This prescription provides maximally  superintegrable Hamiltonians whose spectra presents, as expected,  accidental degeneracy~\cite{quantum}. They are given by  
$$
\begin{aligned}
 {\hH}_{\rm I}&=\frac{1}2  {\hbq^2\left( |\hbq|^{-\bt}+\k |\hbq|^\bt  \right)^2}\, \>\hp^2+A {  \left(|\hbq|^{-\bt}-\k |\hbq|^\bt \right)} ,\\
 \hH_{\rm II}&=\frac{\hbq^2\left( |\hbq|^{-2\ga}-\la^2 |\hbq|^{2\ga}  \right)^2}{2\left( |\hbq|^{-2\ga}+\la^2 |\hbq|^{2\ga}-2\del\right)} \, \>\hp^2 +
\frac{B} { \left(|\hbq|^{-2\ga}+\la^2 |\hbq|^{2\ga}-2\del \right)}   .
 \end{aligned}
$$
\noindent
$\bullet $  {\em Conformal  LB quantization}. The  quantum free Hamiltonian could be  constructed by means of the 
usual LB operator  $\Delta_{\rm LB}$ for the metric $g$ (\ref{aax}):
$$
\begin{aligned}
 {\hH}_{\rm LB, I}&=- \tfrac 12  {\hbar^2}  \Delta_{\rm LB}  +A {  \left(|\hbq|^{-\bt}-\k |\hbq|^\bt \right)}, \\
\hH_{\rm LB,  II}&=-  \tfrac 12  {\hbar^2}  \Delta_{\rm LB}  +
 {B} { \left(|\hbq|^{-2\ga}+\la^2 |\hbq|^{2\ga}-2\del \right)}^{-1}.
 \end{aligned}
$$
However, such quantum Hamiltonians are not maximally superintegrable and their spectrum does not  
convey accidental degeneracy. Therefore these operators define {\em different} physical systems with respect to the above  ones.
Nevertheless, maximal superintegrability can be restored in the LB quantization by considering the so-called conformal LB operator  $\Delta_{\rm CLB}$,
$$
  \Delta_{\rm CLB} = \Delta_{\rm LB} -  \frac{  (N-2)}{4(N-1)} \,R(|\bq|)    ,
  $$
  where $R$ is the scalar curvature (\ref{aac}).
  Then it can be shown that~\cite{quantum}
$$
\begin{aligned}
 {\hH}_{\rm CLB, I}&=- \tfrac 12   {\hbar^2} \Delta_{\rm CLB}  +A {  \left(|\hbq|^{-\bt}-\k |\hbq|^\bt \right)} ,\\
 \hH_{\rm CLB,  II}&=-  \tfrac 12  {\hbar^2}  \Delta_{\rm CLB} +
 {B} { \left(|\hbq|^{-2\ga}+\la^2 |\hbq|^{2\ga}-2\del \right)^{-1}}  ,
 \end{aligned}
$$
are    superintegrable Hamiltonians and present   accidental degeneracy.

 To end with, we stress that the Schr\"odinger and the conformal  LB quantizations are related by means of a  gauge transformation. Thus they have the     same spectrum and differ in the explicit expressions of the eigenfunctions.
In particular, if we consider the Schr\"odinger equations
$\hH \Phi=E\Phi$ and $ \hH_{\rm CLB} \Phi_{\rm CLB}=E\Phi_{\rm CLB}$ for types I and II,   
then~\cite{quantum}
$$
\hH_{\rm CLB}=f( |\hbq|)^{(2-N)/2}\hH f( |\hbq|)^{(N-2)/2},\quad \Phi_{\rm CLB}=f( |\hbq|)^{(2-N)/2}\Phi,
$$
where we have used the conformal factor of the metrics (\ref{zm}) and (\ref{zn}), namely,
$$
 f_{\rm I}( |\hbq|)^2=  \frac 1{\hbq^2\left( |\hbq|^{-\bt}+\k |\hbq|^\bt  \right)^2}   ,\quad  
   f_{\rm II}( |\hbq|)^2=   \frac{\left( |\hbq|^{-2\ga}+\la^2 |\hbq|^{2\ga}-2\del\right)} {\hbq^2\left( |\hbq|^{-2\ga}-\la^2 |\hbq|^{2\ga}  \right)^2} .
\label{ag}
$$


\section*{Acknowledgments}

This work was partially supported by the Spanish MINECO through the Ram\'on y Cajal program (A.E.) and under grants  MTM2010-18556 (A.B and F.J.H.), AIC-D-2011-0711 (MINECO-INFN)  (A.B, F.J.H.~and O.R.) and   FIS201-22566 (A.E.), by  the ICMAT Severo Ochoa under grant SEV-2011-0087 (A.E.),  by Banco Santander-UCM under grant GR35/10-A-910556 (A.E.), and by  a postdoctoral fellowship  by the Laboratory of
Mathematical Physics of the CRM, Universit\'e de Montr\'eal (D.R.).



\end{document}